\documentstyle[floats,twocolumn,aps,prb]{revtex}

\newcounter{three}
\begin{document}
\draft
\title{
\vspace*{-1cm}\hfill
 {\tt }
       \vspace{1cm}\\
{\it Ab initio} study of step formation and self-diffusion on Ag(100)
}
\author{Byung Deok Yu and Matthias Scheffler} 
\address{Fritz-Haber-Institut der Max-Planck-Gesellschaft,
Faradayweg 4-6, D-14195 Berlin-Dahlem, Germany}

\date{\today}

\twocolumn[
\maketitle


\vspace*{-10pt}
\vspace*{-0.7cm}
\begin{quote}
\parbox{16cm}{\small
Using the plane wave pseudopotential method we performed density functional theory 
calculations on the  stability of steps and self-diffusion processes 
on Ag(100). Our calculated step formation energies show that the \{111\}-faceted 
step is more stable than the \{110\}-faceted step. In accordance with experimental
observations we find  that the equilibrium 
island shape should be octagonal very close to a square with predominately
\{111\}-faceted steps. For the (100) surface of
fcc metals atomic migration proceeds by a hopping or an exchange process.
For Ag(100) we find that adatoms diffuse across flat surfaces 
preferentially by hopping.  
Adatoms approaching the close-packed \{111\}-faceted step edges descend from
the upper terrace to the lower level by an atomic exchange with an energy barrier 
almost identical to the diffusion barrier on flat surface regions. 
Thus, within our numerical accuracy ($\approx \pm 0.05$~eV) 
there is no additional step-edge barrier to descent. This provides
a natural explanation for the experimental observations of  
the smooth two-dimensional growth in homoepitaxy of Ag(100).     
Inspection of experimental results of other fcc crystal surfaces indicates that
our result holds quite generally. 

} 
\end{quote} ]

\narrowtext

%
\section{Introduction}
The study of the morphology of surfaces is one of the oldest  
areas of crystal growth and has attracted particular attention
in recent years due to advances in epitaxy and the possibility
of controlled growth of new materials and bulk quality surfaces.
The structure of an epitaxial film is often thought 
to be determined by 
thermodynamic principles, in which the film is assumed to 
adopt the energetically stable thermal-equilibrium structure.~\cite{Bauer} 
However, there are many examples which 
demonstrate that under typical growth conditions 
growth is often governed by kinetics rather than thermodynamics.
A lot of experimental works 
show that the growth mode (the evolution of 
film structure with coverage) can be altered by varying the growth conditions
such as the substrate temperature and deposition rate, or by introducing
defects or surfactants.\cite{Scheff96} 
A notable example is the study of homoepitaxial growth 
of Ag(111): Under clean surface conditions the growing surfaces 
are very rough with mountains as high as
30-40 atomic layers (multilayer growth).\cite{Vrijmoeth,Bromann} 
These studies revealed that
the film structure is governed by the kinetics of surface diffusion.  
Hence, knowledge of the kinetics such as mass transport across
terraces and steps and the diffusion parallel to steps is essential to 
reach understanding of the morphology and
quality of growth. 
   
To date most calculations of atomic diffusion on surfaces employ 
computationally fast semiempirical methods. 
They have been widely used 
by methodological simplicity and under the assumption that the results,
although quantitatively not very accurate, may still
explain general trends.
The main approximation of these methods concerns
the kinetic energy operator of the electron and the neglect of
a self consistent adjustment of the electron density. 
Thus the essence of the mechanism governing the breaking and making 
of chemical bonds is missing.
Obviously, quite often, as {\em e.g.} in the example discussed
in this paper, quantitative results determine the qualitative
features.
For example, when the energy-barrier at the step edge is bigger than
the diffusion barrier at the flat surface the growth mode is rough,
and when the step-edge barrier is equal or smaller, 
the growth mode is flat. The proper treatment
which gives quantitatively and qualitatively more accurate results
and takes the quantum mechanical properties and 
the chemistry of interatomic interactions properly into
account is density functional theory (DFT). 
 
In this paper we report DFT total energy calculations of step formation and diffusion of
Ag adatoms on the (100) surface of fcc silver, 
which extends work presented earlier.\cite{Yu96}

Homoepitaxy of silver exhibits a different behaviour in the growth mode 
depending on substrate orientations: The growing surface of Ag(100) is smooth and
flat whereas that of Ag(111) is typically rather rough.
For Ag(100) surfaces reflection high-energy electron diffraction (RHEED)
intensity oscillations have been
observed~\cite{Suzuki} in the temperature range from 200 to 480 K, 
which indicates that 
the growth proceeds by a smooth two-dimensional (2-D) mode. In contrast, for Ag(111)
surfaces RHEED intensity oscillations are absent at all temperatures, suggesting
that there is no 2-D growth but that the surface becomes very rough
upon growth. This finding for Ag(111) was confirmed by 
x-ray reflection experiments~\cite{Vegt} and 
scanning tunneling microscopy (STM) measurements.~\cite{Vrijmoeth,Bromann}
The former study shows that the growth mode changes from multilayer
to step-flow growth over the temperature range from 175 to 575 K. 
The latter studies
show an STM topography of ``mountain and canyon''-like structures 
with heights of 30-40 atomic layers. 

>From their analysis of STM images for Ag(111)
Vrijmoeth {\em et al.}~\cite{Vrijmoeth} concluded 
that the multilayer growth is attributed to
an additional step-edge barrier. This is  the barrier for 
a diffusing adatom descending steps minus
the surface diffusion barrier. The authors obtained that
the additional energy barrier is
0.15~eV. Bromann {\em et al.}~\cite{Bromann} developed a method to determine 
with high accuracy 
attempt frequencies and activation energies for terrace diffusion as well as 
for the step-down process. This analysis was done by measuring the
nucleation rate on top of
islands as a function of island size and temperature. They obtained for
the additional
energy barrier a value of 0.12~eV very close to 
the 0.15~eV by Vrijmoeth {\em et al.} 
Recent STM studies by Morgenstern {\em et al.}~\cite{Morgenstern}
find the additional step-edge barrier to be 0.10~eV.
This additional
energy barrier which adatoms encounter at step edges explains why the growth
of silver perpendicular to the (111) surface proceeds 
by a very rough multilayer mode. Interestingly, as noted above, the growth of 
the (100) surface is qualitatively different.
As the calculations in Ref.~\onlinecite{Yu96} showed the additional step-edge
barrier is practically zero. Below we extend our previous study and
repeat convergence tests.

\begin{figure}[b]
  \leavevmode
  \includegraphics{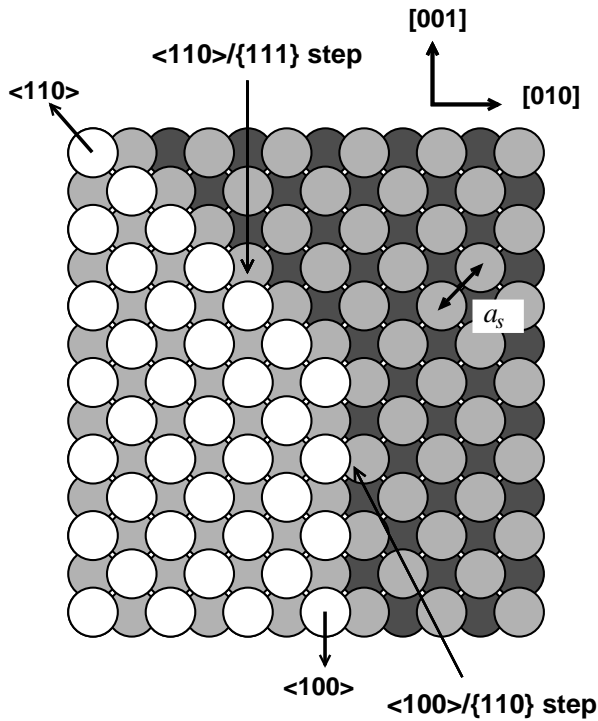}
  \vspace*{10.0cm}
\caption{Top view of monolayer-high steps on a fcc\,(100) terrace:
 a \{111\}-microfaceted step running along 
 the $\langle$\,110\,$\rangle$ direction 
 and a \{110\}-microfaceted step running along 
 the $\langle$\,100\,$\rangle$ direction. 
 The two step edges are labeled 
 as $\langle$\,110\,$\rangle$\,$/$\,\{111\} and
 $\langle$\,100\,$\rangle$\,$/$\,\{110\}, respectively. 
 For our example Ag\,(100) the nearest-neighbor spacing, {\em i.e.}, a surface
 lattice constant $a_s$, is 2.92~\AA.
         }
\label{steps}
\end{figure}
Furthermore,  we consider the shape of islands formed on Ag\,(100) in thermal
equilibrium. There are two types of monolayer-high steps for fcc\,(100) surfaces 
as shown in
Fig.~\ref{steps}: One is a close-packed step running along 
the $\langle$\,110\,$\rangle$ 
direction which has a \{111\} microfacet at edges and the other 
an open step running along the $\langle$\,100\,$\rangle$ 
direction which has a \{110\} microfacet at edges. 
A ratio of formation energies of these two steps determines the equilibrium shape 
of islands on Ag(100) (as obtained by the Wulff construction).~\cite{Zangwill} 
Here we apply the 
{\it ab initio} total energy calculations to the energies of these two sorts of
steps.   

The paper is organized as follows. In Sec.~II, we describe the method of
calculation. In Sec.~III, the equilibrium shape of islands is discussed
as it follows from 
our calculated step-formation energies. In Sec.~IV we present results
for diffusion processes of a Ag adatom on flat and stepped Ag(100) surfaces,  
and discuss their implication on epitaxial growth.
Finally the paper is concluded by Sec.~V.

\section{Method of Calculation}
The total energies and forces on atoms are calculated by 
the DFT. We use a norm-conserving pseudopotential method together with 
the Car-Parrinello-like technique for iterative minimization of the Kohn-Sham total 
energy.~\cite{Stumpf94b} The electronic wavefunctions are expanded in a plane wave 
basis set. Most calculations have been carried out primarily
using the local-density approximation (LDA) 
for the exchange-correlation functional.~\cite{Ceperley}
At important geometries we repeat the calculations using the generalized
gradient approximation (GGA) of Perdew {\em et al.}~\cite{Perdew} 

We use a repeating slab geometry to simulate the actual surface. 
A (3$\times$3) periodicity is employed in the lateral directions,
which we tested gives that artificial adatom-adatom interaction is
sufficiently weak and in fact negligible for the questions of concern.
An artificial periodicity is imposed along the perpendicular direction
to the surface to construct a three-dimensional unit cell. 
We take a slab of three atomic layers and of a 10.35~\AA{} vacuum region.
This rather thin slab is acceptable, as  
we adsorb the adatom only on one side.~\cite{Neugebau92}
We relax the adatom and the top-layer atoms according to a damped Newton 
dynamics, while the other atoms in the two bottom layers are kept at 
the bulk positions. The optimized geometries are identified by
the requirement that the remaining 
forces acting on the atoms are smaller than 0.05~eV/\AA. 
 
The LDA (GGA) ionic pseudopotentials including the tightly bound 4$d$-electron state
as valence states are created using the nonrelativistic (semirelativistic) scheme
of Troullier and Martins.~\cite{Troullier} 
The fully separable norm-conserving pseudopotentials 
of the Kleinman-Bylander form~\cite{KB,Gonze91,Fuchs} are constructed with 
the $s$ pseudopotential as the local component. 
We use a kinetic energy cutoff of 40 Ry to expand the wavefunctions in plane waves. 
The {\bf k}-space integration is performed using nine equidistant
{\bf k} points in the surface Brillouin zone (SBZ) of the (3$\times$3) cell. 
To improve the {\bf k}-space
integration and stabilize convergence 
the electronic states are occupied according to a Fermi distribution 
with $k_{\mbox{\footnotesize B}}T_{\rm el}=$\, 0.1~eV,~\cite{Neugebau92} 
and then the total energies are extrapolated to zero
electronic temperature. In order to attain the iterative solutions of the 
Schr\"odinger equation we start with initial wavefunctions obtained from the 
self-consistent solutions to the Kohn-Sham Hamiltonian in
a mixed basis set of pseudo atomic orbitals and plane waves with a cutoff energy
of 4 Ry, as developed by Kley {\em et al}.~\cite{FHI96MD} This approach  
gives much faster convergence than the conventional 
one starting with the initial wavefunctions created from
the diagonalization of the Hamiltonian and a small cutoff energy in the plane wave 
basis set.

To test the accuracy of the pseudopotentials, we calculated the ground state
properties of bulk silver, using 512 {\bf k} points in the 
Brillouin zone. The results derived from
the Murnaghan equation-of-state fit to the calculated data are
listed in Table~\ref{bulk}. Zero-point vibrations are not considered
in these theoretical results. The calculated values are in good agreement with
$T \rightarrow 0$~K experimental data. 
\narrowtext

\begin{table}[t]
\caption{Structural properties of fcc silver obtained from the 
 plane wave pseudopotential and all-electron linearized-augmented-plane-wave
 (LAPW) methods, for nonrelativistic (NR) and semirelativistic (SR) calculations
 in the LDA and GGA. The $T \rightarrow 0$~K experimental data are 
 given for comparison. 
 }
\begin{tabular} {lccccc} 
 & \multicolumn{2}{c}{This work} & \multicolumn{2}{c}{LAPW\tablenotemark[1]} 
 & Expt.\tablenotemark[2] \\
 & \multicolumn{1}{c}{LDA} & \multicolumn{1}{c}{GGA} 
 & \multicolumn{1}{c}{LDA} & \multicolumn{1}{c}{GGA} & \\
 & \multicolumn{1}{c}{(NR)} & \multicolumn{1}{c}{(SR)} 
 & \multicolumn{1}{c}{(SR)} & \multicolumn{1}{c}{(SR)} & \\
\tableline
$a_0$ (\AA) & 4.14 & 4.18 & 4.00 & 4.17 & 4.07 \\
$B_0$ (GPa) & 99 & 90 & 136 & 85 & 102 \\
\end{tabular}
\tablenotetext[1]{Reference \onlinecite{Khein}}
\tablenotetext[2]{The experimental values are taken 
                  from Reference \onlinecite{Khein}}
\label{bulk}
\end{table}

\section{Step formation on (100) surfaces}
\subsection{Structure and energetics of the flat surface}
Before we describe the step formation on Ag(100) 
it is worthwhile to study the properties of the flat Ag(100) surface.
The surface formation energy per surface atom is given as 
\begin{equation}
 \sigma = (E_{\rm slab} - N_{\rm Ag} \mu_{\rm Ag})/N_{\rm surf}\quad, \label{one}
\end{equation} 
where $E_{\rm slab}$ is the total energy of the slab of $N_{\rm Ag}$ silver atoms 
and $N_{\rm surf}$ is the number of atoms at the surface.  
In thermal equilibrium
the silver chemical potential is equal to the energy of a silver
atom in the bulk.
In order to achieve an optimum of error cancellations
the calculation of the bulk energy $\mu_{\rm Ag}$ is done with 
the same {\bf k}-points sampling as in the slab calculations.
The work function is given by the difference between the electrostatic 
energy in the middle of the vacuum region and the Fermi energy.
The multilayer relaxation $\Delta d_{ij}$ is defined by 
the change in spacing between layers $i$ and $j$
compared to the interlayer spacing in the bulk, $d_0$.

We calculate the properties of flat Ag\,(100), using the slab 
of 5 atomic layers, a vacuum region equivalent
to 5 atomic layers, and 96 {\bf k} points in the 
SBZ of the (1$\times$1) cell. 
In this calculation we used the lattice constant 4.14~\AA{} 
of {\bf k}-sampling-converged calculation which differs only by 0.1~\%{}
from the calculated lattice constant using the above {\bf k}-points.
The results are summarized in Table~\ref{cleanerg}. 
The results in our previous
paper~\cite{Yu96} were obtained using a slab of 4 atomic layers, 
a vacuum region equivalent to 4 layers, and 64 {\bf k}-points in
the SBZ of the (1$\times$1) cell. Due to these changes the present results differ
slightly from those of Ref.~\onlinecite{Yu96}.   
The calculated results are in good agreement with a 
previous full-potential linear-muffin-tin-orbital (FP-LMTO) 
calculation.~\cite{Meth92b,Boisvert} 
Our calculations
of the lattice relaxation give 
$\Delta d_{12}/d_0 = -2.2\%$ and $\Delta d_{23}/d_0 = 0.4\%$.   
The low-energy electron diffraction (LEED)
analysis,~\cite{Li91} in which the first ($d_{12}$) and second ($d_{23}$)
interlayer spacings were determined, 
shows small relaxations with $\Delta d_{12}/d_0 = 0\pm1.5\%$
and $\Delta d_{23}/d_0 = 0\pm1.5\%$. This result
is in good agreement with the present DFT-LDA results
within the experimental accuracy,  
(see also the influence of surface vibrations, which
is neglected here.~\cite{Cho96}).

\narrowtext

\begin{table}[t]
\caption{Calculated results for the properties of flat Ag(100): 
 surface formation energy $\sigma$, work function $\phi$, and 
 multilayer relaxation $\Delta d_{ij}$ relative to the interlayer
 spacing in the bulk $d_0$. 
 It is noted that the experimental surface energy is an estimate 
 from liquid surface tension measurements.
 }
\begin{tabular} {lcccc} 
  & $\sigma$      & $\phi$  & $\Delta d_{12}/d_0$ & $\Delta d_{23}/d_0$ \\
  & (eV/atom)     & (eV)    & (\%)  & (\%)\\ 
\tableline
This work (NR--LDA)  &  0.57   & 4.39 & -2.2  & 0.4   \\
FP-LMTO (NR--LDA)\tablenotemark[1]   &  0.63   & 4.43 & -1.9    &    \\ 
FP-LMTO (SR--LDA)\tablenotemark[2]   &  0.62   & 4.77 & -1.9    &    \\ 
Expt.                   &  0.65\tablenotemark[3] & 4.64\tablenotemark[4] 
 & $0\pm 1.5$\tablenotemark[5] &  $0\pm1.5$\tablenotemark[5]  \\ 
\end{tabular}
\tablenotetext[1]{Reference \onlinecite{Meth92b}}
\tablenotetext[2]{Reference \onlinecite{Boisvert}}
\tablenotetext[3]{Reference \onlinecite{Tyson}}
\tablenotetext[4]{Reference \onlinecite{CRC89}}
\tablenotetext[5]{Reference \onlinecite{Li91}}
\label{cleanerg}
\end{table}
 
\subsection{Structure and energetics of steps}
\begin{figure}[t]
  \leavevmode
  \includegraphics{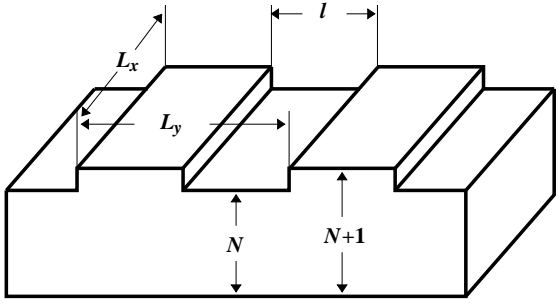}
  \vspace*{5.0cm}
\caption{Schematic representation of the $N$-layer slab of the grooved 
 surface. Periodic boundary conditions are applied along the $x$ and $y$ 
 axes.
         }
\label{grooved}
\end{figure}

In this subsection we consider the stability of the two  
most densely packed monolayer-high steps on Ag(100): 
the close-packed $\langle$\,110\,$\rangle$\,$/$\,\{111\} and
open $\langle$\,100\,$\rangle$\,$/$\,\{110\} steps (see Fig.~\ref{steps}). 
To compute step energies we use the slab geometry of the grooved surface 
sketched in Fig.~\ref{grooved}. 

The simulation cell has the periodicity of $L_x$ 
along the step edge and of $L_y$ along the perpendicular direction 
to the step edge. We take $L_x$ to be $a_s$ and $\sqrt{2}a_s$
for \{111\}- and \{110\}-faceted steps, respectively.
$L_y$ is treated as a variable,
so that the ledge separation $l$ is equal to $L_y/2$ 
(see Fig.~\ref{grooved}). 
The formation energy of the grooved surface $E_{\rm surf}$ is extracted from
the calculations of the $N$-layer slab shown in Fig.~\ref{grooved}, 
by applying the formula
\begin{equation}
  E_{\rm surf}(N) = E_{\rm slab}(N) - N_{\rm Ag} \mu_{\rm Ag}\quad. \label{two}
\end{equation} 
The step formation energy per unit length 
is  
\begin{equation}
  \lambda(N) = (E_{\rm surf}(N) - N_{\rm surf} \sigma(N))/L_x\quad. \label{three}
\end{equation} 
In Eq.~\ref{three}, $\sigma(N)$ is the surface energy per surface atom  
from the calculations of the $N$-layer slab of the flat (100) surface
and $N_{\rm surf}$ is the number of surface atoms at the grooved surface: 
$N_{\rm surf}=L_xL_y/a_s^2$. 
The surface energy
is evaluated by averaging the surface energies of the thin and
thick region of the slab, {\em i.e.} 
 $ \sigma_{\rm ave}(N)=(1-\alpha)\sigma(N)+\alpha\sigma(N+1)$
($\alpha$ is a ratio of the number of surface atoms in the thick region
and the total number of surface atoms). 
The averaging improves the accuracy of the calculated 
step energies. 

Several calculations were done to test the dependence of the calculated 
step energies on the ledge separation $l$ and the number of atomic layers $N$.
The results of the \{111\}-faceted step are summarized in Table~\ref{length}. 
It shows that 
the step energy decreases with increasing
the ledge separation $l$. 
In particular, the step energy $\lambda$
can be expanded with respect to the ledge separation~\cite{Kodiyalam}
\begin{equation}
\lambda=\lambda_0+A/l^2\quad. \label{four}
\end{equation} 
In this equation the first term is the energy of an isolated step,
and the last term gives the effective interaction between steps.
This term includes possible energetic contributions such as
dipole-dipole or elastic interactions.  
For the 3-layer grooved
surface we obtain 
$\lambda_0=0.150$~eV/$a_s$ and $A=0.041$~eV/$a_s^3$ 
in the relaxed geometry
by fitting the variation of the step energy 
with the ledge separation $l$ to Eq.~\ref{four}.  
We find that the step energy already reach the isolated one
at $l=3a_s$ with an error of 0.005~eV/$a_s$. 
The step energy decreases by 0.024~eV/$a_s$
with increasing the slab thickness from 3 to 4 layers.
We find that, 
while the step energy changed by 20~\%{} as the slab thickness is 
increased from 3 to 4 layers, 
a ratio of step energies changes only by 5~\%.
Thus, we see that the slab of 4 layers is sufficient  
to obtain reasonably converged result of a ratio of step energies.
\narrowtext

\begin{table}[t]
\caption{Convergence of the formation energy of the \{111\}-faceted
 step in the relaxed geometry as a function of ledge separation $l$ 
 and the number of atomic layers $N$, calculated by DFT-LDA. 
 }
\begin{tabular} {lcccc} 
$N$ &   3  &  3  &  3 &  4 \\
$l$ ($a_s$)  & 2 & 3 & 4 & 3 \\
\tableline
$E_{surf}$ (eV) [eq.\ref{two}]           &  4.951  &  7.258  &  9.593  &  7.172  \\
$\sigma_{\rm ave}$ (eV/atom)        &  0.579  & 0.579   &  0.581  &  0.576  \\
$\lambda$ (eV/$a_s$) [eq.\ref{three}]     &  0.160  & 0.155   &  0.152  &  0.130  \\ 
\end{tabular}
\label{length}
\end{table}

Using the 4-layer slab of the grooved surface and the ledge separation
of $3a_s$, we obtain the formation energy of 0.130~eV/$a_s$ (LDA) for
the close-packed \{111\}-faceted step. For the \{110\}-faceted step we 
take a 4-layer slab and
ledge separation of $2\sqrt{2}a_s$.
The obtained step energy is 0.156~eV/$a_s$ (LDA).
Thus, the close-packed 
\{111\}-faceted step is more stable than
the open \{110\}-faceted one.    
This confirms the experimental observation that the 
\{111\}-faceted steps are preferentially formed in thermal equilibrium.
This result is not unexpected 
because of atomic geometries, {\em i.e.} the local coordinations of
the two steps differ noticeably: Step-edge atom at the \{111\}-faceted step
has seven nearest 
neighbors, while the step-edge atom at the \{110\}-faceted step has only 
six neighbors. 
Using the AFW (Adams-Foiles-Wolfer~\cite{Adams}) 
EAM (embedded atom method) functions, 
Nelson {\em et al.}~\cite{Nelson} found $\lambda$ to be 0.102~eV/$a_s$ and
0.142~eV/$a_s$ for \{111\}- and \{110\}-faceted steps, 
respectively. With the VC (Voter-Chen~\cite{Voter}) functions they obtained 
$\lambda$ to be 0.135~eV/$a_s$ for \{111\}-faceted step.  
\begin{figure}[b]
  \leavevmode
  \includegraphics{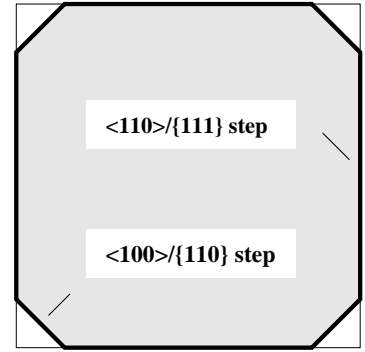}
  \vspace*{5.5cm}
\caption{Calculated shape of islands in thermal equilibrium.
 The close-packed \{111\}-faceted edges are longer with an edge length ratio
 $L^{\langle 100 \rangle /\{110\}}$\,:\,$L^{\langle 110 \rangle /\{111\}}$ 
 of 3:10 than the open \{110\}-faceted ones. 
         }
\label{shape}
\end{figure}

When the formation energies of all steps were known, the equilibrium
shape of an island can be obtained by finding the island shape with minimum
free energy, or by applying the Wulff construction.~\cite{Zangwill} 
For the (100) surface it is certainly plausible that only two steps are
important.
Thus we expect
octagonally shaped islands as shown in
Fig.~\ref{shape}. The calculated step formation energies 
imply that
the \{111\}-faceted edges should dominate and that
the edge-length ratio should be 
$L^{\langle 100 \rangle /\{110\}}$\,:\,$L^{\langle 110 \rangle /\{111\}}=3$\,:\,10.
This theoretical finding can be compared to experimental
observations.
In fact, for ion-bombarded Ag(100) surface the preferential
formation of the close-packed \{111\}-faceted step edges has
been observed at room temperature,~\cite{Teichert}
and for Ir\,(100) Chen and Tsong, using the field ion microscope, 
also found that the equilibrium island shape is rectangular with 
\{111\}-faceted steps.~\cite{Chen95}

\begin{figure}[b]
  \leavevmode
  \includegraphics{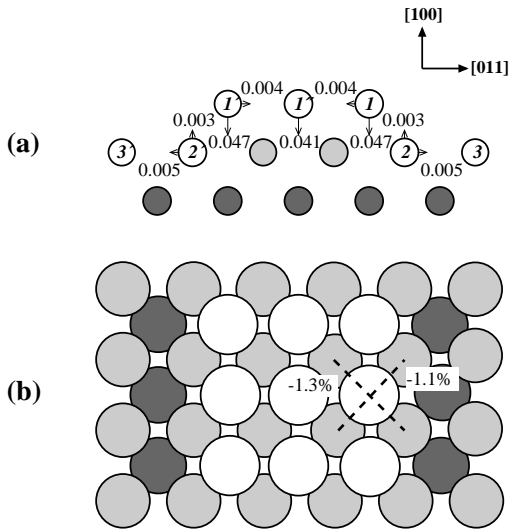}
  \vspace*{7.7cm}
\caption{(a) Side and (b) top views of stable geometry of 
 the grooved surface of the ledge separation 3$a_s$ 
 with \{111\}-faceted steps. In (a) arrows and numbers 
 in unit of \AA{} indicate directions and magnitudes of atomic displacements 
 from the unrelaxed ideal step geometry. The percentage changes in distances
 to nearest-neighbors of step-edge atoms  
 relative to the nearest-neighbor separation in bulk are shown in (b). 
         }
\label{geometry}
\end{figure}
Next we consider the relaxations of the stepped surface. Figure~\ref{geometry}
shows the geometry optimized structure of \{111\}-faceted steps.
The relaxations of surface atoms in the upper terrace  
are nearly identical to those of surface atoms of flat Ag(100): surface 
atoms 1, $1^{\prime}$, and $1^{\prime\prime}$ relax toward the bulk 
by $\sim$ 2\% of the bulk interlayer
spacing $d_0$ [see Fig.~\ref{geometry} (a)]. 
Atoms 2 and $2^{\prime}$ at step bottom slightly relax 
upward and toward atoms 3 and $3^{\prime}$, but
these relaxations of 
lower terrace atoms are nearly negligible. Figure~\ref{geometry} (b)
displays the bond lengths of step-edge atoms
with their underlying nearest neighbors and
the numbers show that the bond lengths are very close to those 
between surface atoms and
their nearest neighbors in the subsurface sites of flat Ag\,(100). 
Similarly, at the grooved surface with the \{110\}-faceted steps
downward relaxations of upper terrace atoms dominate.

\begin{figure}[b]
  \leavevmode
  \includegraphics{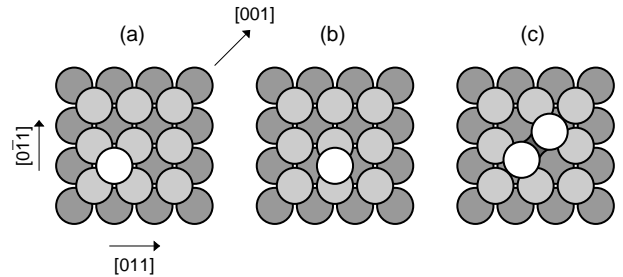}
  \vspace*{3.5cm}
\caption{Top view of an adatom at (a) the fourfold hollow site,
 (b) the transition state for the hopping diffusion (the twofold bridge site), 
 and (c) the transition state for the exchange diffusion. 
 The white and gray circles represent the ad- and surface atoms, respectively.
         }
\label{geo}
\end{figure}

\section{Self-diffusion on flat and stepped (100) surfaces}
\subsection{Adsorption and self-diffusion on flat surfaces}
We now present the results of adsorption and 
diffusion of a Ag adatom on the flat  Ag\,(100) surface. 
For metals, metal adatoms on surfaces preferentially adsorb at the 
highest coordination site. 
Indeed the equilibrium adsorption site for a Ag adatom on Ag\,(100) is 
the fourfold hollow [Fig.~\ref{geo}(a)].  
In the optimized geometry, 
the four nearest neighbors of the adatom 
slightly distort in the lateral directions, opening the hollow site even further.
The lateral displacement from the ideal clean surface site is 0.02~\AA.
The adatom is located 1.84~\AA{} above the surface layer,
and the bond length between the adatom and its neighbors is
2.78~\AA{}, {\em i.e.}, 5~\%{} shorter than the interatomic distance in the
bulk. 
This follows the typical trend, namely that bond strength
per bond decreases with coordination and correspondingly, bond length
increases with coordination. 

In metal-on-metal surface diffusion, an adatom typically
moves across a flat surface by a series of hops between
adjacent equilibrium adsorption sites.  
For fcc\,(100) an adatom in a fourfold site 
moves over the twofold bridge site
into a neighboring fourfold equilibrium site [see Fig.~\ref{geo}(b)].
An adatom may visit all of the fourfold surface sites,
forming a ($1\times1$) square pattern.
An alternative mechanism for surface diffusion is atomic exchange where
a diffusing atom moves by displacing a neighboring
surface atom. For the fcc\,(100) surface the exchange process
of an adatom with a surface atom occurs preferentially in the [\,010\,] 
and [\,001\,] directions. At the saddle point of the transition
the adatom and surface atom are above a vacated surface-layer site  
[see Fig.~\ref{geo}(c)].
Subsequently the displaced surface atom becomes a new adatom in 
a next-nearest-neighbor
fourfold site. 
When this diffusion mechanism is active,
an adatom may visit only half of the fourfold surface sites,
forming a c($2\times2$) square pattern. 
Experimental evidences of the self-diffusion by exchange displacement on 
fcc\,(100) 
have been obtained by Kellogg and Feibelman on Pt\,(100)~\cite{Kellogg} 
and by Chen and Tsong on Ir\,(100)~\cite{Chen90}. 
The process has been 
theoretically studied by Feibelman for Al\,(100).~\cite{Feibelman}
His total-energy calculations of the self-diffusion barrier on Al\,(100)
show that atomic diffusion via exchange is energetically favored by a factor
of three over hopping.  
It is an open question if the exchange diffusion plays a role for other
fcc\,(100) surfaces as well.
The EAM calculations of Liu {\em et al.}~\cite{Liu} suggest that, 
besides Pt(100) and Ir(100), exchange displacements are also favored
for self-diffusion on the (100) surfaces of Pd and Au.
For self-diffusion on Cu(100) effective medium calculations~\cite{Hansen}
predicted that exchange displacement occurs preferentially,
but this result is 
in contrast to other semiempirical studies~\cite{Liu} and 
to a DFT-LDA calculation.~\cite{Lee94}

Our total-energy calculations for the self-diffusion on Ag\,(100) 
clearly show that the hopping mechanism is energetically favored.
The energy barriers for hopping diffusion are
0.52~eV (LDA) and  0.45~eV (GGA) which is close to the result
of the FP-LMTO LDA calculation~\cite{Boisvert},
which got 0.50~eV. Using the AFW and VC EAM functions 
Liu {\em et al.}~\cite{Liu} obtained
a similar value, namely 0.48~eV.
Our results show that the use of the GGA functional for the exchange-correlation
energy lowers the energy barrier by 13~\%{} (0.07~eV) compared to the LDA result. 
Recent STM studies of homoepitaxy on Ag(100)~\cite{Zhang96}, 
combined with Monte-Carlo simulations
of an appropriate model for fcc\,(100) homoepitaxy, determined the diffusion barrier 
to be 0.33~eV. Low-energy ion scattering measurements of 
Langelaar {\em et al.}~\cite{Lang96}
obtain a value of 0.40~eV.
At this point we note that the experimental analyses of the diffusion
barriers are not without problem.
Thus, we conclude that our GGA result ($E_d = 0.45$~eV)~\cite{Yu96} is well
in agreement with (subsequently obtained) available experimental
information.
In the optimized geometry of the bridge site, 
the saddle point for the hopping diffusion, 
the two nearest neighbors of the adatom are pushed away 
and downwards by 0.05~\AA. The adatom is located 2.33~\AA{} above the 
surface layer. The bond length between the adatom and its two 
neighbors is 2.69~\AA{}, {\em i.e.} 3 \%{} shorter than in the fourfold hollow.
This again follows the well known trend:
Each of the two bonds at the bridge site is stronger than each of the
four bonds at the hollow site.
\begin{figure}[t]
  \leavevmode
  \includegraphics{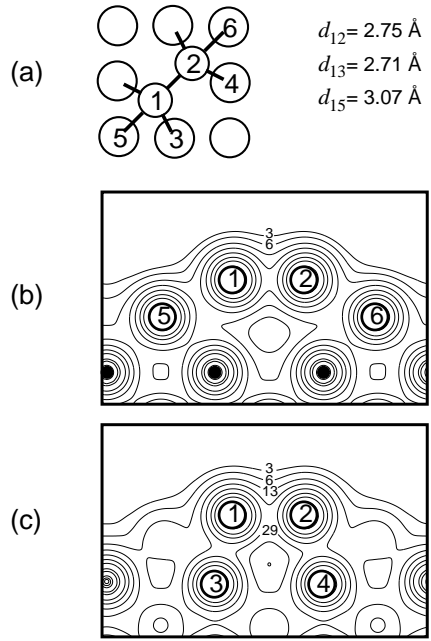}
  \vspace*{9.6cm}
\caption{Valence charge density at the transition state for the 
 exchange diffusion.
 A schematic plot of the optimized atomic geometry is shown in (a). 
 Panels (b) and (c) display the electron density.
 In (b) the plot is in the plane cutting atoms 5, 1, 2, and 6.
 In (c) the plot is in the plane cutting atoms 3, 1, 2, and 4.
 The value of the lowest-density contour is 
 3$\times 10^{-3}$ $e$/({\rm bohr})$^{3}$ and
 subsequent contours differ by a factor of 2.2. Solid circles mark 
 the second layer atoms.
         }
\label{charge}
\end{figure}

To obtain the energy barrier for the exchange diffusion we compute
the total energy of the transition state shown in Fig.~\ref{geo}(c).
We obtain an energy  barrier of 0.93~eV (LDA) and 0.73~eV (GGA) 
which is much higher than that for the hopping process. Thus, we see that 
Ag adatoms diffuse across flat regions of Ag\,(100) by hopping.
It is interesting that for the exchange geometry the difference
between LDA and GGA is noticeable (0.20~eV). 
The GGA value is comparable to the EAM results.
Using the AFW and VC EAM functions Liu {\em et al.} found
an energy barrier of 0.75~eV and 0.60~eV for the exchange process,
respectively.
In the optimized geometry of the transition state for the exchange diffusion
[see Fig.~\ref{charge}(a)], each of the two top-most adatoms has four bonds with
neighbors. The distance between the two top-most adatoms is
2.75~\AA, 1 \%{} shorter than in the fourfold hollow.
These two atoms are located 1.52~\AA{} above the surface layer. 
The four atoms neighboring these atoms are pushed away by 0.06~\AA{}
and downwards by 0.05~\AA. The bond length between the top-most and
underlying Ag atoms is 2.71~\AA, 3~\%{} shorter than in the fourfold hollow.
Another bond length between the atoms 1 and 5 (between the atoms 2 and 6) in 
Fig.~\ref{charge}(a) is 3.07~\AA{}, 10~\%{} longer than in the fourfold hollow,
and 5~\%{} longer than in the bulk.
We note that at the (100) surface of aluminum, the exchange displacement is
supported by the formation of covalent bonds, which is made possible
(and indeed plausible) by the $sp$ valence of the group  
{\Roman{three}} element Al.
For silver we find that the
electron density does not reflect a pronounced covalency effect
as shown in Fig.~\ref{charge}(b) and (c), 
which explains why the exchange displacement does not play a role 
at flat Ag\,(100).
  
\subsection{Adsorption and self-diffusion on stepped surfaces}
\begin{figure}[t]
  \leavevmode
  \includegraphics{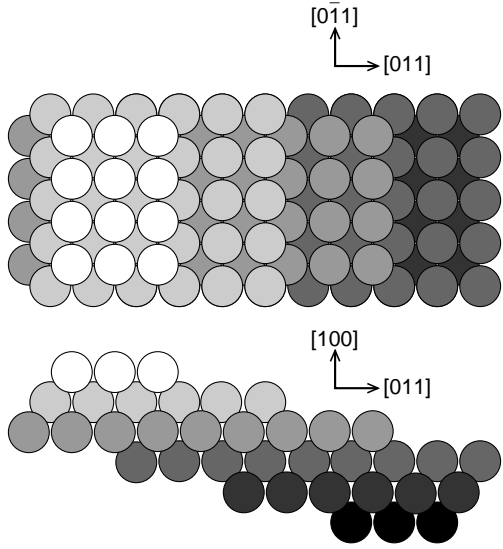}
  \vspace*{8.0cm}
\caption{Top and side view of the fcc\,(511) surface. The (511) surface 
 has close-packed \{111\}-faceted steps and the number of atomic rows 
 within the (100)-oriented terraces is three. 
         }
\label{vicinal}
\end{figure}
\begin{figure}[t]
  \leavevmode
  \includegraphics{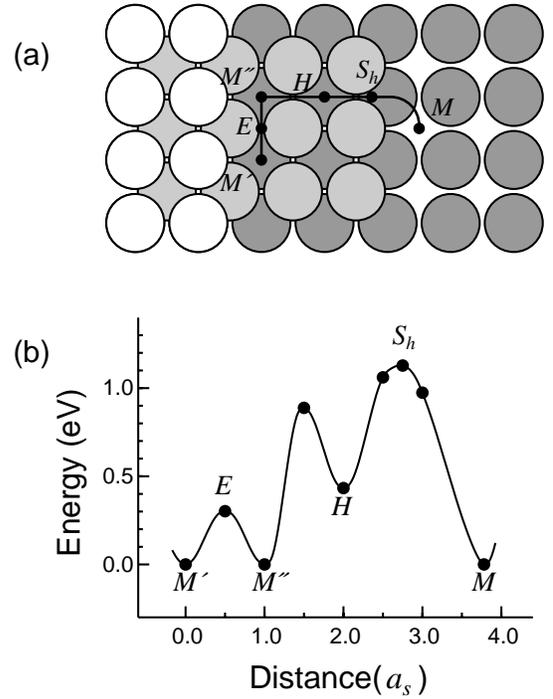}
  \vspace*{10.5cm}
\caption{Total energy of a Ag adatom diffusing along the 
         indicated path by a hopping (roll-over) process, 
         calculated within the LDA.
         A top view of the vicinal (511) surface is shown in (a).
         The \{111\}-faceted edges are aligned along the [0\={1}1] direction.
         The distance is given in unit of the surface lattice constant 
         $a_s$.
         }
\label{hopping}
\end{figure}

This subsection is concerned with the
diffusion of a Ag adatom at the \{111\}-faceted steps
which are preferentially formed on Ag\,(100) surfaces. 
The calculations were done for a high index (511) surface which is vicinal to (100)
(see Fig.~\ref{vicinal}).
This surface consists of (100) terraces which are three atom rows wide.
The step edges are perpendicular to the [100] and the [011] directions.
The periodicity along the step edge is taken to be
three surface lattice constants.
Using a vicinal surface allows to consider a 
smaller number of atoms within the unit cell
than that required for the ordered step arrays on flat surfaces 
(the grooved surfaces).
Using the vicinal (511) surface of 4 atomic layers, 
we obtain the formation energy of 0.136~eV/$a_s$ (LDA) for the close-packed
\{111\}-faceted step, which  
is close to the corresponding value 0.130~eV/$a_s$ (LDA)
from the 4-layer slab calculations of grooved surface (see Table~\ref{length}). 
In Fig.~\ref{hopping}(b) we display the total energy  
of a Ag adatom diffusing
on the stepped Ag\,(100) by the hopping process. For any position of the 
path indicated in Fig.~\ref{hopping}(a) the total energy is obtained by
relaxing all coordinates of the atoms of the surface layer as well as 
the height of the diffusing atom.

Figure~\ref{hopping}(b) clearly shows   
two stable adsorption sites: The hollow site ($M$) at the step, at which
the adatom is fivefold coordinated and
the hollow site ($H$) on the terrace, at which the adatom coordination is four.
Due to the higher coordination it is indeed plausible that the DFT results give
that the adsorption site $M$ is more stable than site $H$.
The energy difference between the $M$ and $H$ sites is 0.43~eV (LDA)
and  0.32~eV (GGA). 
In the optimized geometry of the site $M$ the distance between the step-edge
atom and adatom is 2.89~\AA{}, 1 \%{} shorter than in the bulk.
The bond length between the two step-bottom atoms and adatom is
2.78~\AA, 
and that between the two lower-terrace atoms and adatom is 2.84~\AA.
It is also interesting to note that desorption of an adatom at
the step edge to the flat region requires to overcome a very high energy
barrier (LDA: 0.96~eV; GGA: 0.76~eV).

When the adatom rolls over the ledge from $H$ site on the upper terrace 
to $M$ site on the lower terrace, we obtain an energy barrier of 
0.70~eV (LDA) and 0.55~eV (GGA). Thus, we find the ``additional
step-edge barrier'' to be 0.18~eV (LDA) and 0.10~eV (GGA)
to step down from the upper terrace by the hopping (roll-over) process. 
This value is by about
30~\% (20~\%) higher in the LDA (GGA) than the diffusion barrier at the flat region.
At the transition state for the roll-over process the adatom is 
located near the bridge site at the ledge [$S_h$ in Fig.~\ref{hopping}(a)].
In the optimized geometry the two nearest neighbors of the adatom are
slightly pushed away, and the adatom is located 2.27~\AA{} above the 
upper terrace.
The bond length between the step-edge atoms and adatom is found to be 2.72~\AA{},
1 \%{} longer than in the bridge site at the flat region
and 7 \%{} shorter than that of a bulk atom: Each of the two bonds
at site $S_h$
is weaker than that at the bridge site on the flat region.
At the bridge site on the flat terraces the adatom has four 
next-nearest-neighbors,
while at $S_h$ the adatom has two next-nearest-neighbors. 
Thus, we could understand the additional energy barrier in terms
of weaker bonds between the adatom and underlying surface atoms, and
decrease of the number of the next-nearest-neighbors 
compared to the flat surfaces.  

\begin{figure}[t]
  \leavevmode
  \includegraphics{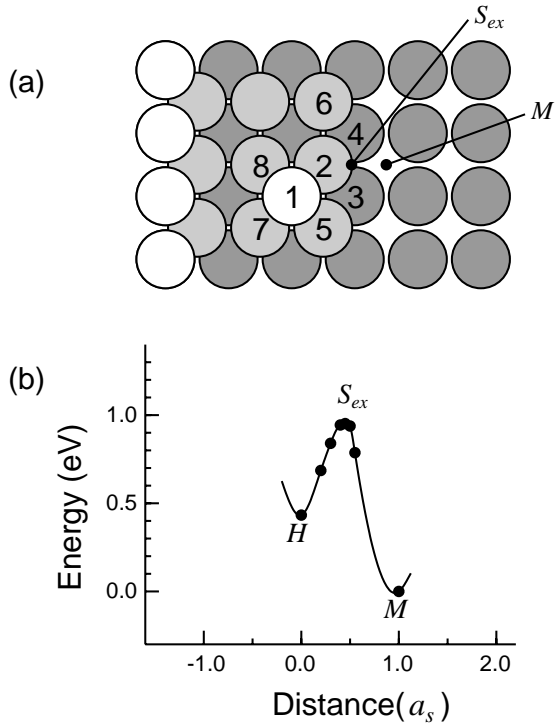}
  \vspace*{10.5cm}
\caption{Total energy of a Ag adatom diffusing across a step by an
         exchange process, 
         calculated within the LDA.
         In (a) the adatom 1 is adsorbed at the fourfold hollow site $H$.
         The total energy as a function of  
         the distance of the step-edge atom 2 from the undistorted step edge
         is shown in (b). 
         }
\label{exchange}
\end{figure}
\begin{figure}[t]
  \leavevmode
  \includegraphics{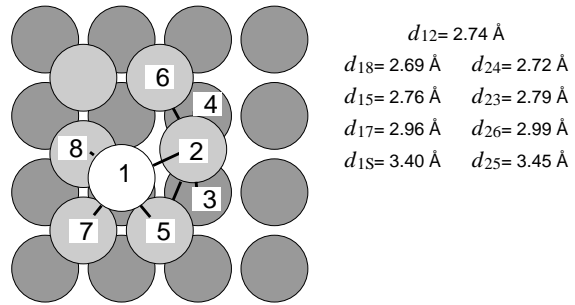}
  \vspace*{5.0cm}
\caption{Schematic view of the optimized atomic structure at the
         transition state $S_{ex}$ of the step-down diffusion by the
         exchange process, 
         calculated within the LDA.
         $d_{1{\rm S}}$ denotes the bond length between the adatom 1
         and second-layer atom below.
         }
\label{s_ex}
\end{figure}

The other possibility to step down from the upper terrace 
is via an exchange, where adatom 1 displaces step-edge atom 2 and subsequently 
the latter moves to the fivefold coordinated site $M$ [see Fig.~\ref{exchange}(a)].
Figure~\ref{exchange}(b) displays the results of the step-down diffusion
by the exchange process.
Within our numerical accuracy for energy differences ($\approx \pm 0.05$~eV)
the calculated energy barrier of 0.52~eV (LDA) and 0.45~eV (GGA) is almost identical 
to that of the hopping diffusion at the flat region.
The transition state for the exchange process is identified to be
near the bridge site
formed by two step-bottom atoms 3 and 4 on the lower terrace
[$S_{ex}$ in Fig.~\ref{exchange}(a)].
The geometry of the optimized structure of the transition state
is given in Fig.~\ref{s_ex}.
The top-most adatom (no.~1) is located 1.26~\AA{} above the upper-terrace atoms.
The height is much lower than the value of 1.52~\AA{} at the flat region. 
Inspection of the geometry at $S_{ex}$ shows that
the local coordination of atoms 1 and 2 concerned with the exchange displacement
remains high, since the atom 1 follows in close contact to atom 2 during
the process: At $S_{ex}$ each of the two atoms 1 and
2 has five bonds with neighbors, 
while at the saddle point of the exchange diffusion on the flat region 
each of the two top-most atoms has only four bonds [Fig.~\ref{geo}(c)].
The origin of the lower diffusion barrier of the step-down motion by the
exchange displacement compared to the flat region 
is thus the additional bonds formed at the obtained saddle point.  
Our finding that {\it there is no additional energy barrier to
descend from the upper to the lower terrace}
provides a natural explanation for the
smooth 2-D growth of silver in the (100) orientation.
By analyzing the STM images of Ag(100) during deposition
Zhang {\em et al.}~\cite{Zhang96} gave an estimate of $0.025\pm0.005$~eV 
for the additional step-edge barrier. 
Within our numerical accuracy ($\approx \pm 0.05$~eV) 
this is in good agreement with 
our result that the additional step-edge barrier is negligible. 

Our finding that for Ag(100) an adatom 
goes down at step edges by an exchange process is expected to
apply for other noble metals and other fcc transition metals as well.
In fact, for homoepitaxial growth of Cu(100),~\cite{Ernst92} 
Pd(100),~\cite{Flynn93} and Fe(100)~\cite{Stroscio93}
damped oscillations of the scattering intensity   
have been observed.  
In order to explain the intensity oscillations associated with smooth
2-D layer-by-layer growth, the step-edge barrier for the descent
of adatoms was assumed to be negligible.
For a hopping (roll-over) process we expect that an adatom generally 
encounters an
additional (positive) step-edge barrier [$\sim 0.1$~eV for Ag\,(100)]
because of the low coordination of the transition state. 
Thus, we expect that quite generally for fcc\,(100) metal surfaces 
step-down motion
proceeds by an exchange process, similarly to Ag(100).

Finally we address the atomic motion along the close-packed 
\{111\}-faceted step to get ideas about the possible roughness of steps
and kinetic growth shapes of islands.
We find that the diffusion along step edge proceeds by a hopping process.
An adatom in a fivefold coordinated adsorption site moves over
a bridge site $E$ into a neighboring fivefold coordinated site
[see Fig~\ref{hopping}(a)].
>From Fig.~\ref{hopping}(b) we see
that the barrier (LDA: 0.30~eV; GGA: 0.27~eV) is significantly
lower than the surface diffusion barrier $E_d$ (LDA: 0.52~eV; GGA: 0.45~eV). 
This lower energy barrier indicates that 
Ag islands formed on Ag\,(100) should be 
compact. Atoms which reach the step edges will certainly be able to diffuse parallel
to the steps and thus local thermal equilibrium is attained. 
We therefore expect rather straight step edges and 
no fractally shaped islands. 
In the optimized geometry of site $E$ the distance between the two step-edge
atoms and adatom is 2.99~\AA, 2 \%{} longer than in the bulk.
We find the bond lengths of 2.66~\AA{} between the step-bottom atom and adatom
and of 2.84~\AA{} between the lower terrace atom and adatom
and note that the coordination varies from five to four as
the adatom moves from site $M^{\prime}$ to site $E$ [Fig.~\ref{hopping}(a)], 
while for diffusion on flat terraces the coordination
varies from four to two.   
Thus, the lower barrier for diffusion parallel to steps can be understood 
in terms of smaller variation in coordination.

\section{Conclusion}
In the above Section, we presented results of 
first-principles total-energy 
calculations for the electronic structure and energies of steps 
and for various microscopic self-diffusion 
processes at Ag\,(100). 

For fcc\,(100) surfaces two types of monolayer-high steps are of particular
importance: the close-packed
\{111\}- and more open \{110\}-faceted steps.
Our total-energy calculations show that the \{111\}-faceted step
is more stable than the \{110\}-faceted one. 
This is understood in terms of different coordination of step-edge
atoms. In accordance with experimental observations, we find that
in thermal equilibrium  the shape of island should be octagonal
but very close to a square with \{111\}-faceted edges.
The edge-length ratio of the octagon is calculated as 10:3.
The energy barrier for self-diffusion of a Ag adatom along \{111\}-faceted 
edges is found to be
significantly lower than the surface diffusion barrier. 
Thus, we expect rather straight step edges and no fractally shaped
islands.  

At flat regions of Ag\,(100)  
Ag adatoms are found to diffuse by a hopping process.
The obtained energy barrier is 0.45~eV in the GGA which
is very close to the experimental estimates which are 0.40~eV\cite{Lang96} 
and 0.33~eV~\cite{Zhang96}.
In contrast to atomic motion on flat region,
the descent of adatoms at steps proceeds by an exchange process.
The calculated energy barrier is almost identical to the barrier at flat region.
This indicates that there is 
no additional step-edge 
barrier to diffuse across step edges. This finding is in sharp contrast to 
the additional energy barrier 
($\Delta E_{\rm step}^{\rm Ag(111)}=0.15$~eV~\cite{Vrijmoeth} and
0.12~eV~\cite{Bromann} and 0.10~eV~\cite{Morgenstern}) experimentally found at  
Ag\,(111) actuating a rough growth morphology.
The calculated result implies good interlayer mass transport at Ag\,(100) and
thus explains the smooth 2-D growth experimentally observed in homoepitaxy of
Ag\,(100). Inspection of experimental works indicates that the step-down motion
at steps by an atomic exchange takes place quite generally for
other fcc metals.

We thank A.~Kley, S.~Narashimhan, G.~Boisvert, C.~Ratsch, and P.~Ruggerone
for helpful discussions.
Enlightening discussions with G. Rosenfeld on various growth and self-diffusion
experiments are gratefully acknowledged.
B.~D.~Yu gratefully acknowledges a fellowship from the Alexander von Humboldt
Foundation. 


\end{document}